# FIELD-DEPENDENT SURFACE RESISTANCE OF A SUPERCONDUCTING RF CAVITY CAUSED BY SURFACE IMPURITY

M. Ge[#], F. Furuta, M, Liepe, G.H. Hoffstaetter
Cornell University, Ithaca, NY, 14853, USA

*Abstract*

Q-slope issue, which is caused by the field dependent surface resistance, puzzled people for a long time in SRF fields. In this paper, we related the Q-slope with surface treatments; and proposed a surface-impurity model to explain the field-dependent of surface resistance of SRF cavities. Eighteen cavity-test results have been analysed to examine the model. These cavities were treated by different recipes: Nitrogen-doping; BCP and HF-rinsing; EP with 120°C baking; and EP without 120°C baking. The performance of these cavities, which is normally represented by cavity quality factor versus accelerating gradient or surface magnetic field curves ($Q_0$ vs. $E_{acc}$ or $Q_0$ vs. B), has included all types of Q-slope, such as Low-field Q-slope, Medium-field Q-slope, and Anti-Q-slope. The data fittings are quite successful; the fitting results were shown. The model can be used to evaluate the effectiveness of the surface treatments. At last, the paper discussed the way to build a high-Q high-gradient SRF cavity.

## INTRODUCTION

The surface resistance ($R_s$) of superconductor under radio-frequency (RF) fields is a critical topic, because it determines the intrinsic quality-factor ($Q_0$) of SRF cavities by inverse proportionality. People explored many effective methods to reduce the surface resistance at an accelerating-gradient to obtain the highest Q-value of a SRF cavity. The BCS theory assumes that the surface resistance is non-magnetic-field dependent in a pure and uniform superconductor [1-4]. In real measurements, however, the quality factor has been observed having strong field-dependent effects, which is called Q-slope. These Q-slopes, shown in Fig 1, relate to the recipes of surface preparation, and can be categorised into Low-field Q-slope (LFQS), Medium-field Q-slop (MFQS), and Anti-Q-slope [5]. In this paper, we do not discuss high-field Q-slope case, which is caused by localized hot-spots or global heating [15].

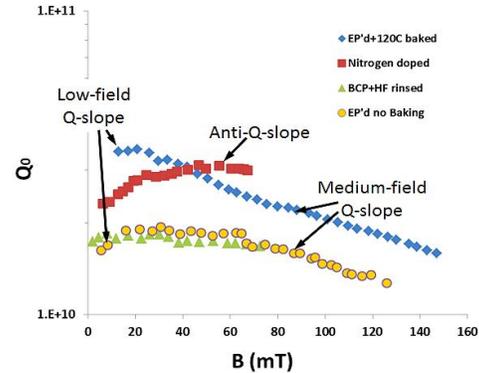

Fig. 1: The Q-slopes after different surface treatments.

One of the standard recipes of a cavity treatment is electro-polishing (EP) followed by 120°C baking; its $Q_0$ vs. B curve is shown in Fig.1. (◆). This curve has positive slope (rising $Q_0$) below ~20mT (defined as LFQS), but then the slope changed to negative (dropping $Q_0$) from 20mT to about 80-120mT, defined as MFQS. While the nitrogen-doped cavity [6-8], shown in Fig.1 (■), has positive Q-slope from the low field to medium field (0-80mT), which is called Anti-Q-slope. The buffered chemical polished (BCP) cavity followed by HF acid rinsing has a very flat $Q_0$ vs. B curve [9], shown in Fig.1 (▲). The EP without 120°C baking cavity has similar $Q_0$ vs. B curve with the baked one, shown in Fig.1 (●). All these results manifest that the final finishing layer formed during the surface treatments dramatically affects the superconducting performance of cavities.

The Q-slope issue puzzles SRF people for a long time. Many models and theories have been developed to explain the Q-slope [5, 10-14]. But if those theories and models treated the surface as clean and uniformed, it can explain part of the Q-slope results, but contradict to the others, because the surface layer is formed differently by surface preparations but not uniformed.

In this paper, we propose the surface-impurity model based on a non-uniform and dirty surface as a new approach to explain all types of Q-slope. We also explain that the impurity layer on top of surface in essence causes the Q-slope. Hence, in this proposal, the BCS theory is no need to be modified. We examined this model by 18 cavity-test results, which have been treated by most commonly used recipes: Nitrogen-doping; BCP and HF-rinsing; EP with 120°C baking; and EP without 120°C baking. The fitting results are quite successful and will be shown and discussed in the following sections.

___________________
#mg574@cornell.edu

## IMPURITY EFFECTS

The surface resistance of a superconductor ($R_s$) consists of two parts [16], one is the BCS resistance ($R_{BCS}$) and the other is the residual resistance ($R_0$). It can be written in Eq. (1).

$$R_s = R_{BCS}(T, T_c, \lambda_L, \xi_0, l_e, \Delta) + R_0, \quad (1)$$

here the $R_{BCS}$ is the function of the surface temperature T; and it is also determined by the critical temperature $T_c$, the London penetration depth $\lambda_L$, the coherence length $\xi_0$, the electron mean free path (MFP) $l_e$, and the energy gap (EnGap) $\Delta$. Among these parameters, the MFP $l_e$ represents the surface impurity. When the surface is 'dirty', the mean free path is short, and vice versa.

The dirty on the surface is the layer formed after the surface treatments. It could be oxide layer, nitrogen and Nb compounds, *etc*. In reference [14], it shows the nitrogen content versus depth after the nitrogen doping. These nitrogen and oxygen contents which are called the 'defects' in this paper can serve as scattering centers for superconducting electrons. Fig.2 displays the surface-impurity model. The dirty layer formed underneath the Nb surface; here we only consider the non-uniformity along depth. As the magnetic field applied on the surface, it decays exponentially to zero, described in Eq. (2)

$$B(x) = B(0)e^{-\frac{x}{\lambda_L}}, \quad (2)$$

here $x$ is the depth from RF surface. The London penetration depth $\lambda_L$ is given in Eq. (3):

$$\lambda_L = \sqrt{\frac{m}{\mu_0 n_s q^2}}, \quad (3)$$

where $m$ is the superconducting electron mass; $q$ is the electron charge; $n_s$ is the superconducting electron density, which is related to the exterior magnetic field, i.e. when $B \uparrow$, the density $n_s \downarrow$, hence the $\lambda_L \uparrow$, and $R_{BCS} \uparrow$. However, some cavity test results, e.g. Fig.1 (▲), show very flat $Q_0$ versus $E_{acc}$ curves which implies the $R_{BCS}$ is not sensitive to the $\Delta \lambda_L$. It is also confirmed by the calculation of the BCS theory, shown in Fig. 2. If the $\lambda_L$ varies from 330 to 400Å, the $R_{BCS}$ only increased about $1n\Omega$. In this paper, we neglect the $\Delta R_{BCS}$ directly introduced by the $\Delta \lambda_L$.

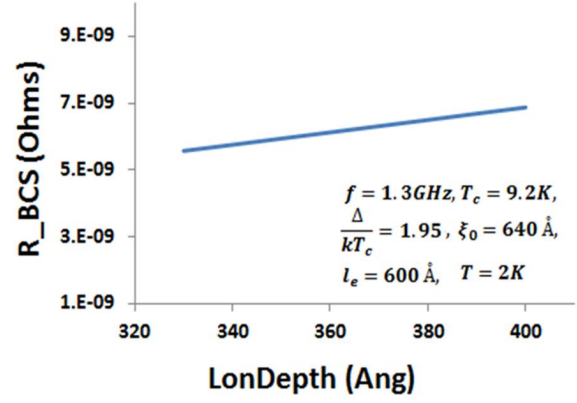

Fig. 2: The BCS resistance versus London penetration depth calculated by the BCS theory.

But the $\Delta \lambda_L$ affects the B-field decay curve in Eq. (2); and leads the $R_{BCS}$ becoming strong field-dependent when the surface has an impurity layer. $B(0)$ in Eq. (2) is the surface magnetic field, which is proportional to the accelerating gradient ($E_{acc}$). If $B(0)$ value is larger, the London penetration depth $\lambda_L$ is longer. Thus $B(0)$ decays slowly and can penetrate into the material deeper, and vice versa. The $\Delta \lambda_L$ causes the changes of the portion of the electrons in the dirty and clean region at different B-field. In Fig. 3, $B_2 > B_1$, hence $\lambda_{L2} > \lambda_{L1}$. The $B_1$ mostly decays in the dirty layer; while for the $B_2$, it can penetrates more into the clean region.

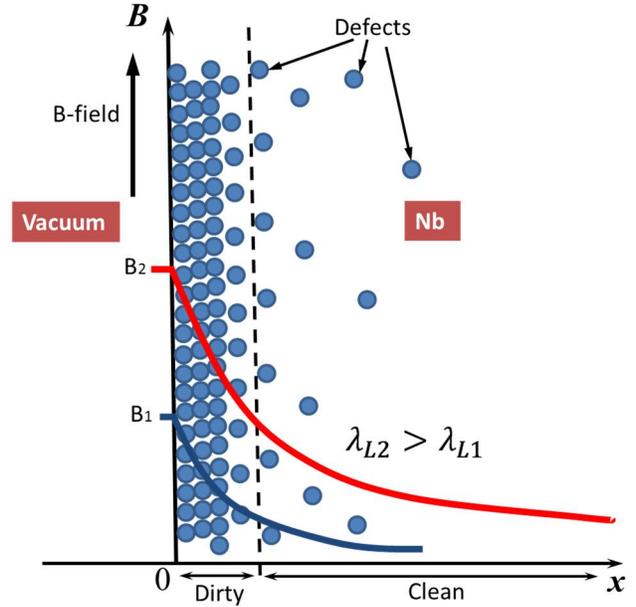

Fig.3: The surface-impurity model scheme.

## The BCS resistance, $R_{BCS}$

To calculate the $R_{BCS}$ for $B_1$ and $B_2$, the mean free path $l_1$ and $l_2$ can be evaluated in Eq. (4),

$$l_1 \approx l_d, \qquad l_2 = f(l_d, l_c), \quad (4)$$

where the $l_d$, $l_c$ denote the mean free path of the dirty and clean layer. Here $l_c > l_d$. When the B-field goes lower as $B_1$, the $l_1$ will be close to the $l_d$; while the B-field goes higher as $B_2$, the $l_2$ has contribution from both of the $l_d$ and $l_c$. As the $B_2$ goes higher, the $l_2$ will be closer to the $l_c$. Thus the mean free path is a function of the magnetic field in a non-uniform and dirty surface. The mean free path value should start from the small value $l_d$ at low field and saturate to the larger value $l_c$ at higher field. We use an error function to describe it in Eq. (5),

$$l_e = l_c \, \text{erf}(\alpha \, B^\beta), \quad (5)$$

here $\alpha$ and $\beta$ are the constants obtained from the data fitting of cavity measurements, which describing the shape of the $l_e$ vs. B curve. They are determined by the dirty surface character, i.e., the defects density distribution along the depth. In the Eq. (5), we do not build $l_d$ directly, because it can be described when B~0mT. The calculated curves of $l_e$ vs. B at different $\alpha$ and $\beta$ are plotted in Fig. 4. This indicates how thin or thick the dirty layer on the surface is. The orange curve ($\alpha=0.05$, $\beta=1.5$) in Fig 4, for example, shows the surface has the thinnest dirty layer compared with the other conditions.

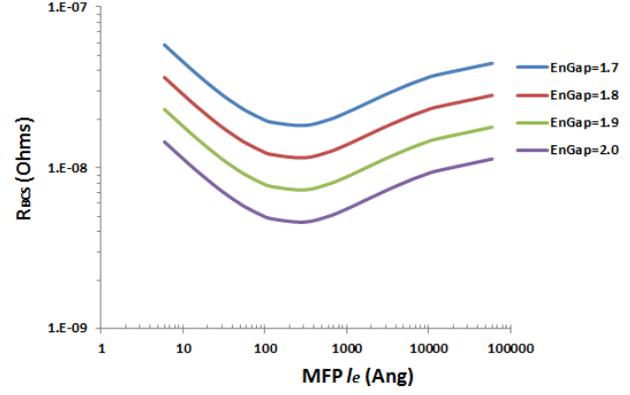

Fig. 5: The BCS resistance versus MFP in Energy Gap range from 1.7-2.0 at frequency 1.3GHz and temperature 2K calculated by SRIMP code.

Fig. 5 clearly shows that the BCS resistance has the minimum value when the mean free path is about 200~400Å. When $l_e < \xi_0$, the $R_{BCS}$ decreases as the mean free path increases; it has been called the 'dirty RF limits'. When $l_e > \xi_0$, which is the 'clean RF limits', the $R_{BCS}$ shows anomalous behavior, increasing with the mean free path.

The changes of the mean free path $l_e$ will cause the coherence length $\xi$ change described in Eq. (6),

$$\frac{1}{\xi} = \frac{1}{\xi_0} + \frac{1}{l_e}, \quad (6)$$

where $\xi_0$ denotes the coherence length in the clean and uniform material at temperature 0K.

## The residual resistance, $R_0$

Since the dirty layer not only affects the superconducting electrons but also affects the behavior of the normal conducting electrons, *i.e.* the dirty layer will dominate the residual resistance value at low field. As the magnetic field becomes larger and penetrates into deeper clean region, the contribution of dirty layer could be relatively small and the clean layer will dominate the contribution on the residual resistance. It will result in the reduction of $R_0$. Therefore the relation of residual resistance $R_0$ versus magnetic field can be constructed in quite similar way, described in Eq. (7)

$$R_0 = R_a - R_b \, \text{erf}(\gamma \, B), \quad (7)$$

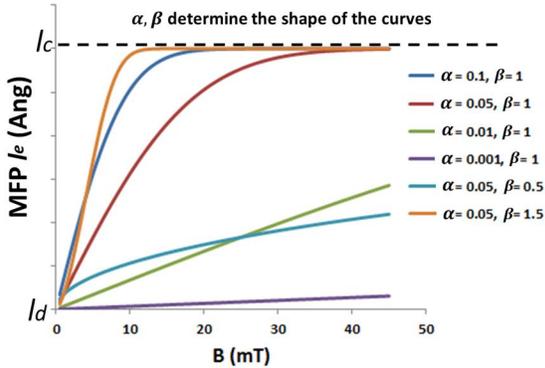

Fig. 4: The mean free path vs. magnetic field curves described in Eq. (5).

Fig. 5 plots the $R_{BCS}$ vs. the mean free path in different Energy Gap at frequency 1.3GHz and temperature 2K. The curves were calculated by the BCS theory (SRIMP code).

here $R_a$, $R_b$ and $\gamma$ are constants fitted from measurement data. The physical meaning of Eq. (7) is addressed by Fig. 6 which gives the calculated curves of Eq. (7). $R_a$ represents the value in the dirty layer when the field is low; $R_a$-$R_b$ is the value in the clean region; and the residual resistance saturates at $R_a$-$R_b$ when the field is strong. $\gamma$ describes the shape of the curve.

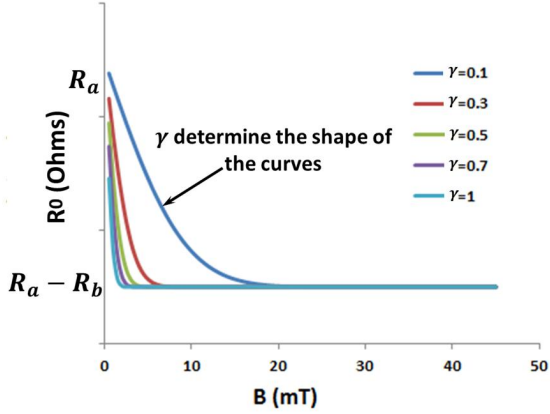

Fig. 6: $R_0$ vs. magnetic field curves described in Eq. (5).

*The surface resistance, $R_s$*

Eq. (1), Eq. (5), Eq. (7) can be combined together to obtain the field dependent surface resistance $R_s$, in Eq. (8),

$$R_s(T, B) = R_{BCS}[T, T_c, \lambda_L, \xi_0, l_c \, \text{erf}(\alpha B^\beta), \Delta] + [(R_a - R_b \, \text{erf}(\gamma B))]. \quad (8)$$

SRIMP code has been adopted to calculate the BCS resistance [17, 18]. SRIMP code, written by J. Halbritter, incorporates the BCS theory. The temperature $T$ in Eq. (8) is the temperature on inner surface of a cavity, but not a bath temperature. It can be calculated based on the thermal feedback model [19, 20] from the bath temperature and the accelerating gradient [21].

## MEASUREMENT DATA ANALYSIS

*The fitting method*

The calculated surface resistance can be converted into Q-value by geometry factor $G$ in Eq. (9),

$$R_s = \frac{G}{Q_0}. \quad (9)$$

For a given cavity shape, the ratio of the surface peak magnetic field $B_p$ and $E_{acc}$ is a constant. Then Eq. (9) can be written as,

$$Q_{0\_calc}(T_{bath}, E_{acc}) = \frac{G}{R_s[T(T_{bath}, E_{acc}), E_{acc}]}. \quad (10)$$

In the cavity tests, we measured a series of $Q_0$ values at different gradients and bath temperatures described in Eq. (11),

$$Q_{0\_meas.}(T_{bath}, E_{acc}) = \begin{bmatrix} Q_{00} & \cdots & Q_{0m} \\ \vdots & \ddots & \vdots \\ Q_{n0} & \cdots & Q_{nm} \end{bmatrix} \begin{matrix} @T_0 \\ \vdots \\ @T_n \end{matrix} \quad (11)$$
$$@E_0 \quad \cdots \quad @E_m$$

The fitting takes every $Q_0(T_{bath})$ curves at different $E_{acc}$; compares Eq. (10) and Eq. (11); and tunes the parameters $\alpha, \beta, \gamma, R_a, R_b, l_c, \Delta$ to achieve the minimum fitting error by the least squares method. The fitting error *RSS* is given by Eq. (12):

$$RSS = \sum_{\substack{i=1 \\ j=1}}^{n,m} [Q_{0\_calc.}^{i,j}(T_{bath}^i, E_{acc}^j) - Q_{0\_meas.}^{i,j}(T_{bath}^i, E_{acc}^j)]^2. \quad (12)$$

*The fitting results*

The vertical test (VT) data are obtained from 1.3GHz Nb fine grain multi-cell and single-cell cavities. The bath temperature range was from 1.6K to 2K. All the data were taken without quench and field emission. In the fitting, we set $T_c = 9.2K, \lambda_L = 360$Å$, \xi_0 = 640$Å [5]. Fig. 7 gives the plot of four $Q_0$ vs. $E_{acc}$ curves and their fitting results at temperature 2K. The cavities were treated by different recipes. They are listed as follow,

o  (♦): *EP 80μm, 120°C baking,*
o  (■): *EP 120μm, Nitrogen-doped, light EP 26μm,*
o  (●): *EP 38μm,*
o  (▲): *BCP 135μm, 120°C baking, HF rinsed.*

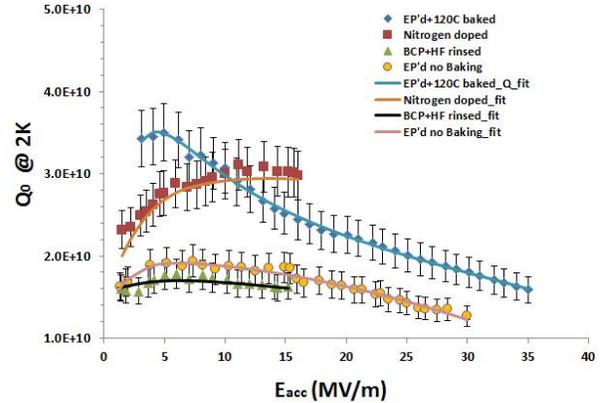

Fig.7: The $Q_0$ versus $E_{acc}$ curves of different treatments recipes and their fitting results at temperature 2K, 10% errors of the measurements were given.

The fitting by the surface-impurity model agrees with the test data very well. All types of Q-slope can be fitted very well. The fitted parameters are listed in Table 1,

Table 1: The fitted parameters obtained from the $Q_0$ vs. $E_{acc}$ curves in Fig. 7.

| | α | β | γ | $R_a(10^{-9})$ (Ohms) | $R_b(10^{-9})$ (Ohms) | $l_c$ (Ang) | $\frac{\Delta}{kT_c}$ |
|---|---|---|---|---|---|---|---|
| EP, 120C baking | 0.0024 | 1.81 | 1.17 | 3.04 | 2.26 | 6000 | 1.95 |
| $N_2$-doped | 0.0018 | 0.68 | 0.25 | 4.36 | 2.17 | 11100 | 1.92 |
| EP | 0.0127 | 0.46 | 0.32 | 3.95 | 3.07 | 6120 | 1.74 |
| BCP, HF rinsed | 0.0026 | 0.15 | 0.21 | 6.83 | 1.73 | 78000 | 1.82 |

## DISUCCIONS

The fitted parameters in Table 1 can be put into Eq. (5), Eq. (7), and Eq. (8), it's possible to obtain the mean free path vs. $E_{acc}$, $R_{BCS}$ vs. $E_{acc}$, and $R_0$ vs. $E_{acc}$ curves. They are plotted in Fig. 8, Fig. 9, Fig. 10 and Fig. 11.

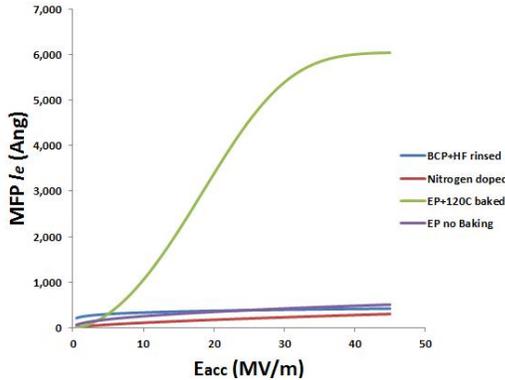

Fig. 8: The mean free path versus $E_{acc}$ curves of different treatments, which is calculated by Eq. (4) from the fitting parameters in Table 1.

The curve of EP and 120°C baking increases with $E_{acc}$ rapidly compared to the other three curves. It suggests that the dirty layer formed by EP and 120°C baking is thinner than the other three. The 120°C baking diffused the defects into deeper material, thus the defect density along the depth has been reduced.

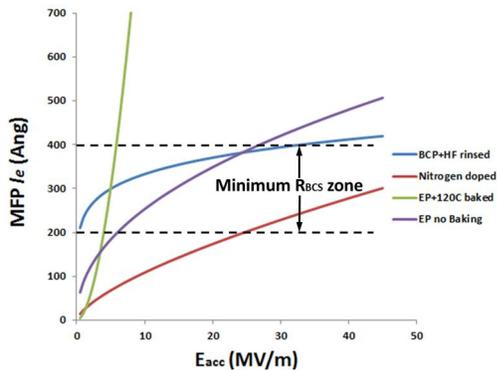

Fig. 9: The fine scale of Fig. 7 to show the minimum BCS resistance zone.

Fig. 9 shows the finer scale of Fig. 8 from 0 to 700Å. It allows showing better view of the other three treatments. As it has been described in the previous section and shown in Fig 5, the $R_{BCS}$ has the minimum zone when the mean free path is in between 200 to 400Å. Here we can regard the 400 Å line as the boundary of the dirty limits and the clean limits. When the mean free path value passes it $R_{BCS}$ begin to increase (Fig. 4). The mean free path curve of EP and 120°C baking passed the boundary around 5MV/m and kept increasing, hence the $R_{BCS}$ decreased up to 5MV/m (LFQS), and then it increased, which caused the MFQS above 5MV/m. The nitrogen-doped cavity curve stays always below the boundary up to $E_{acc}$ of ~25MV/m, which implies the $R_{BCS}$ is always decreasing (rising $Q_0$), which is the reason of the Anti-Q-Slope below 25MV/m. The mean free path curve of BCP and HF rinsed cavity stays in the minimum zone, which explains why its $Q_0$ vs. $E_{acc}$ curve is flat below 30MV/m. While the MFP of EP without baking entered the zone before 5MV/m (LFQS); then stays in the zone. This is consistent with the $Q_0$ curve is relatively flat between 5-20MV/m, and then starts to drop after passing the boundary.

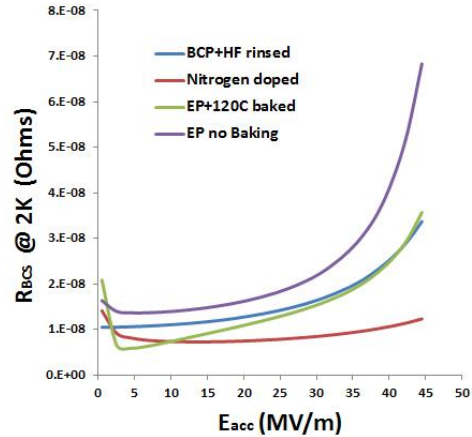

Fig. 10: The BCS resistance versus $E_{acc}$ curves of different treatment at temperature 2K.

The MFP curves determine the shape of the $R_{BCS}$ curve. But the absolute value of the $R_{BCS}$ is also determined by the energy gap. From Fig. 5, we know that the larger energy-gap will give the smaller $R_{BCS}$. The results in Table 1 shows that the EP with 120°C baking and the Nitrogen-doped cavity has larger energy gap value around 1.95. Therefore their BCS resistance is quite small. In Fig. 10, the $R_{BCS}$ of EP without baking increased dramatically above 30MV/m, which is purely caused by the thermal feedback effect. The energy gap of EP without baking is very small (1.74); hence it gave large $R_{BCS}$ which heated up the cavity inner-wall significantly above 30MV/m.

The detail calculations have been addressed in reference [21].

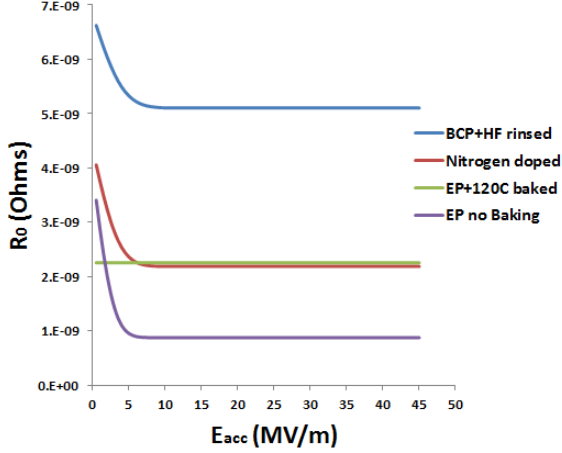

Fig. 11: The residual resistance versus $E_{acc}$ curves of different treatment at temperature 2K.

Fig. 11 shows $R_0$ has no contribution on the field dependent of $Q_0$ above 5MV/m for the recipes we applied on cavities. Especially; the $R_0$ becomes almost constant after 120°C baking on EP cavity. Therefore the LFQS of EP with 120°C baking case is mainly caused by the $R_{BCS}$ changes. And also it shows evidence again that the dirty layer becomes very thin after 120°C baking. The other three treatments have the contributions from both of $R_{BCS}$ (about 5 nΩ changes) and $R_0$ (about 2 nΩ changes) below ~5MV/m. The changes of $R_0$ are smaller compared to the changes of the $R_{BCS}$ at low field in Fig. 10.

The comparison between EP with and without 120°C baking shows that the unbaked cavity has higher BCS resistance but lower residual resistance. In the other words, 120°C baking caused the residual resistance increasing, but the BCS resistance decreasing. In total the surface resistance has been decreased about $4.2 n\Omega$ at 15MV/m by 120°C baking, thus the baked cavity has higher $Q_0$ than the unbaked cavity [16].

## THE CONCEPT OF DESIGN A HIGH-Q CAVITY AT AN ACCELERATING GRADIENT

People are always interested in obtaining a high-Q cavity at a target gradient for a real accelerator. Based on the surface impurity model analysis, after the surface treatment the mean free path should be controlled between the minimum zone ($l_e \sim$200-400Å) as well as the energy gap should be kept as large as possible ($\frac{\Delta}{kT_c} \sim$1.95-2) to achieve the minimum BCS resistance at the target gradient. For the residual resistance, it should be decayed rapidly at the low field and the field dependent of $R_0$ is negligible between middle to high field. Therefore the surface resistance $R_s$ will be minimized by (1) the well-controlled minimum $R_{BCS}$ and (2) rapidly decayed and minimized $R_0$ at the target gradient. The high Q at the target gradient will be obtained as shown in Fig. 12.

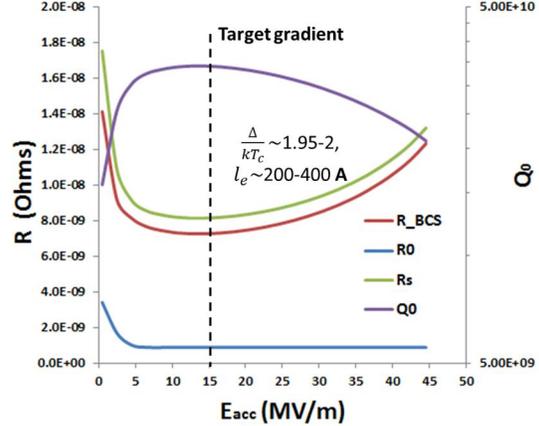

Fig. 12: The concept of design a high-Q cavity at a target gradient.

Nitrogen-doped cavity achieved the highest Q-value (3-5×$10^{10}$) [8] at temperature 2K accelerating gradient 15-16MV/m in the current existing surface treatment method, so far. Table 2 is the statistics of the VT of SRF cavities treated by the four recipes at Cornell University. Nitrogen-doped cavities have the Q-value about $3.0 \times 10^{10}$, because the energy gap and the mean free path are between the ranges in which the BCS resistance is the minimum.

Table 2: The statistic of the measurements at $E_{acc}$=15-16MV/m.

| | N of Tests | $Q_0$ ($10^{10}$) | $\frac{\Delta}{kT_c}$ | $l_e$ (Ang) | $R_0(10^{-9})$ (Ohms) | $R_{BCS}(10^{-9})$ (Ohms) |
|---|---|---|---|---|---|---|
| N$_2$-doped | 8 | 3.0±0.2 | 1.94±0.03 | 142±71 | 2.6±0.5 | 6.7±0.7 |
| EP, 120C baking | 4 | 2.1±0.3 | 1.95±0.04 | 2025±566 | 4.4±0.4 | 9.0±1.9 |
| EP | 2 | 1.6±0.2 | 1.78±0.05 | 665±509 | 2.9**4 ± 2.9** | 14±0.5 |
| BCP, HF rinsed | 4 | 1.75±0.1 | 1.81±0.02 | 202±122 | 3.5±1.1 | 12.4±0.9 |

## CONCLUSION

The surface-impurity model can be used to fit and explain the all types of Q-slope. The data fitting of the model against the measurement data are quite successful. The parameters extracted from different surface treatments can be used to evaluate the effectiveness of the treatments. The method of producing the high-Q cavity at a target gradient has been pointed out.

## REFERENCES

[1] J. Bardeen, L.N. Cooper, J.R. Schrieffer, *Phys. Rev., 108(1957), p. 1175.*

[2] D.C. Mattis and J. Bardeen, *Phys. Rev., 111, 412 (1958).*

[3] A.A. Abrikosov, L.P. Gor'kov, and I.M. Khalatnikov, *Sov. Phys. JETP*, 8, 182 (1959).

[4] J.P. Turneaure, I. Weissman, *Appl. Phys., Vol 39, Number 9.*

[5] H. Padamsee, *RF Superconductivity, Vol. II Science, Technology, and Applications*, WILE-VCH, 2009.

[6] A. Grassellino, A. Romanenko, O. Melnychuk, Y. Trenikhina, A. Crawford, A. Rowe, M. Wong, D. Sergatskov, T. Khabiboulline, F. Barkov, Nitrogen and argon doping of niobium for superconducting radio frequency cavities: a pathway to highly efficient accelerating structures, Supercond. Sci. Technol. **26**, 102001 (2013).

[7] D. Gonnella, R. Eichhorn, F. Furuta, M. Ge, D. Hall, V. Ho, G. Hoffstaetter, M. Liepe), T. O'Connell, S. Posen, P. Quigley, J. Sears, V. Veshcherevich, A. Grassellino, A. Romanenko and D. A. Sergatskov, Nitrogen-doped 9-cell cavity performance in a test cryomodule for LCLS-II, J. Appl. Phys. 117, 023908 (2015).

[8] D. Gonnella, M. Liepe. New Insights into Heat Treatment of SRF Cavities in a Low-pressure Nitrogen Atmosphere. Proceedings of IPAC2014, Dresden, Germany.

[9] A. Romanenko, J. P. Ozelis1, A. Grassellino, Depth distribution of losses in superconducting Niobium, Proceedings of IPAC2012, New Orleans, Louisiana, USA, WEPPC116, P2495.

[10] R.G. Eichhorn, B. Bullock, J.V. Conway, B. Elmore, F. Furuta, Y. He, G.H. Hoffstaetter, J.J. Kaufman, M. Liepe, T.I. O'Connel, P. Quigley, D.M. Sabol, J. Sears, E.N. Smith, V. Veshcherevich. Cornell's Main Linac Cryomodule for the Energy Recovery Linac Project. Proceedings of IPAC2014, Dresden, Germany.

[11] W. Weingarten, *Physical review ST-AB, 14, 101002 (2011).*

[12] B.P. Xiao, C.E. Reece, M.J. Kelley, Superconducting surface impedance under radiofrequency field, Physica C: Superconductivity, 2013. 490(0): p. 26-31.

[13] A. Gurevich, Reduction of Dissipative Nonlinear Conductivity of Superconductors by Static and Microwave Magnetic Fields, Phys. Rev. Lett. 113, 087001 (2014).

[14] R. Eichhorn, D. Gonnella, G. Hoffstaetter, M. Liepe and W. Weingarten, On superconducting niobium accelerating cavities fired under $N_2$-gas exposure, http://arxiv.org/abs/1407.3220.

[15] M. Ge, G. Hoffstaetter, F. Furuta, E. Smith, M. Liepe, S. Posen, H. Padamsee, D. Hartill, X. Mi, A temperature-mapping system for multi-cell SRF accelerating cavities. Proceedings of IPAC2013, Pasadena, CA USA.

[16] A. Romanenko, A. Grassellino, Dependence of the microwave surface resistance of superconducting niobium on the magnitude of the rf field, Appl.Phys.Lett. 102 (2013) 252603.

[17] J. Halbritter, Zeitschrift für Physik 238 (1970) 466.

[18] J. Halbritter, "Fortran Program for the Computation of the Surface Impedance of Superconductors", Internal Notes KFZ Karlsruhe 3/70-6. June 1970.

[19] J. Vines, Y. Xie, H. Padamsee, "Systematic Trends for the Medium Field Q-Slope", Proceedings of SRF 2007, 2007, TUP27, p.178 Beijing, China.

[20] H. Padamsee. Calculations for breakdown induced by "large defects" in superconducting niobium cavities. CERN/EF/RF82-5, 1982.

[21] M. Ge, G. Hoffstaetter, H. Padamsee, V. Shemelin, Extracting superconducting parameters from surface resistivity by using inside temperature of SRF cavities, arXiv:1405.4226 (http://arxiv.org/abs/1405.4226)